\documentclass[twocolumn,prl,superscriptaddress,aps,amsmath,floatfix]{revtex4-2}

\usepackage[normalem]{ulem}

\usepackage{graphics}
\usepackage{graphicx}
\usepackage{epstopdf}
\usepackage{dcolumn}
\usepackage{bm}
\usepackage{longtable}
\usepackage{epsfig}
\usepackage{times}
\usepackage{url}
\usepackage{color}
\usepackage{braket}

\begin{document}
\title{Inhomogeneous mass trap for dark-state polaritons  in  atomic media}

\author{Ding-An \surname{Chen}}
\affiliation{Department of Physics, National Central University, Taoyuan City 32001, Taiwan}

\author{Kai-You \surname{Huang}}
\affiliation{Department of Physics, National Central University, Taoyuan City 32001, Taiwan}

\author{Chun-Yen \surname{Hsu}}
\affiliation{Department of Physics, National Central University, Taoyuan City 32001, Taiwan}
\affiliation{Department of Physics, National Tsing Hua University, Hsinchu 30013, Taiwan}
\affiliation{National Center for Excellence in Quantum Information Science and Engineering, National Tsing Hua University, Hsinchu 30013, Taiwan}

\author{Meng-Cheng  \surname{Xie}}
\affiliation{Department of Physics, National Tsing Hua University, Hsinchu 30013, Taiwan}
\affiliation{National Center for Excellence in Quantum Information Science and Engineering, National Tsing Hua University, Hsinchu 30013, Taiwan}

\author{Ite A. \surname{Yu}}
\affiliation{Department of Physics, National Tsing Hua University, Hsinchu 30013, Taiwan}
\affiliation{National Center for Excellence in Quantum Information Science and Engineering, National Tsing Hua University, Hsinchu 30013, Taiwan}

\author{Wen-Te \surname{Liao}}
\email{wente.liao@g.ncu.edu.tw}
\affiliation{Department of Physics, National Central University, Taoyuan City 32001, Taiwan}
\affiliation{Physics Division, National Center for Theoretical Sciences, Taipei 10617, Taiwan}
\affiliation{Quantum Technology Center, National Central University, Taoyuan City 32001, Taiwan}
\affiliation{National Center for Excellence in Quantum Information Science and Engineering, National Tsing Hua University, Hsinchu 30013, Taiwan}
\date{\today}
\begin{abstract}
The generation of a  trapping potential for dark-state polaritons in a two-dimensional electromagnetically induced transparency system is theoretically studied. We show that such a trap can arise from a spatially inhomogeneous effective mass of the dark-state polariton. Because this mass inhomogeneity can be engineered by tuning the parameters of the control fields, the motion, spatial profile, and coherent behavior of bound dark-state polaritons can be tailored accordingly. Our results enable spatial controls of optical information and provide a possible route toward realizing Bose-Einstein condensation of dark-state polaritons in a trapping potential.
\end{abstract}
\maketitle
Electromagnetically induced transparency (EIT) has revolutionized quantum optics by enabling the coherent manipulation of light-matter interactions \cite{Harris1997}. One of its most striking effects is slow light \cite{Hau1999}, where the group velocity of a probe pulse can be dramatically reduced via controllable dispersion. This mechanism allows reversible mapping between photons and long-lived atomic coherences by switching the control field, forming the basis of optical quantum memory \cite{Phillips2001, Kocharovskaya2001, Chen2013, Hsiao2018, Wang2019, Chu2025}.  Using counter-propagating control fields, stationary light pulses (SLPs) can be created, forming dark-state polaritons (DSPs) \cite{fleischhauer2000} with vanishing group velocity \cite{bajcsy2003, Lin2009, Kim2018, Kim2022, Kim2023}. In this regime, DSPs behave as massive quasiparticles whose effective mass and kinetic properties can be engineered via EIT parameters \cite{Zimmer2008, Fleischhauer2008, Otterbach2010, Kim2025}, enabling long interaction times for quantum information processing and Rydberg-DSP interactions \cite{Fleischhauer2008, Chen2012, noh2017, Kim2021}. However, most studies of SLPs assume spatially uniform control fields \cite{bajcsy2003, Zimmer2008, Fleischhauer2008, Otterbach2009, Lin2009, Liao2009, Otterbach2010, Kim2025}. Although this approximation captures essential features of SLPs, it leaves a vast landscape of unexplored physics of SLP associated with spatially structured control fields \cite{Ling1998, Brown2005, Zimmer2006}. In particular, nonuniform control-field amplitudes can modify the effective mass, dispersion, and potential landscape experienced by DSPs, enabling synthetic gauge fields \cite{Kuan2023} and coherent coupling between DSP spatial modes \cite{Shia2025}. Here, we show that shaping counter-propagating control fields with tailored amplitude and phase structures generates scalar and vector  potentials for DSPs. The resulting complex potential provides a coherent confinement through its real part and spatially dependent attenuation through its imaginary part. Bound DSP states emerge from the interplay between these components. 
Our results pave the way for a realization of the Bose–Einstein condensation of DSPs \cite{Fleischhauer2008, Kim2021} in a trapping potential \cite{Cornell2002, Pethick2008}.

\begin{figure}[b]
\includegraphics[width=0.48\textwidth]{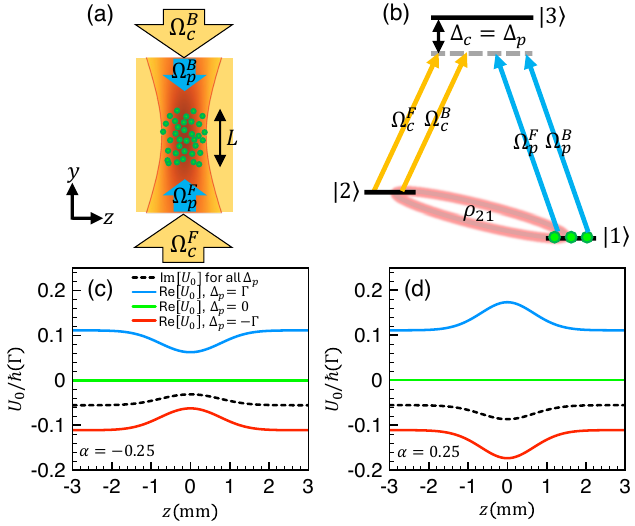}
\caption{\label{fig1}
(a) Two-dimensional counter-propagating EIT system. Green dots represent the atomic medium of length $L$. Yellow arrows denote the counter-propagating biased-Gaussian control fields $\Omega_c^{F(B)}$, and blue arrows indicate the probe fields $\Omega_p^{F(B)}$.
(b) Three-level $\Lambda$-type scheme, where $\Omega_p^{F(B)}$ and $\Omega_c^{F(B)}$ drive the transitions $\ket{1} \rightarrow \ket{3}$ and $\ket{2} \rightarrow \ket{3}$ with detuning $\Delta_p = \Delta_c$.
Inhomogeneous mass trap potential $U_0/\hbar$ with (c) $\alpha = -0.25$ and (d) $\alpha = 0.25$.
Red-solid, green-solid, and blue-solid line depict the real part of $U_0/\hbar$ for $\Delta_p=-\Gamma$, $0$, and $\Gamma$, respectively. Black-dashed line illustrates the imaginary part of $U_0/\hbar$. Other parameters are  $\left( w_0, \phi, \xi\right) = \left( 1 \text{ mm}, 0, 80\right) $.
}
\end{figure}
We investigate a two-dimensional, counter-propagating, three-level $\Lambda$-type electromagnetically induced transparency (EIT) system \cite{Harris1997, Harris1990, fleischhauer2000, Liu2001, Phillips2001, Kocharovskaya2001}, as illustrated in Figs.~\ref{fig1}(a) and (b).
The control  fields (yellow  arrows) couple the $\ket{2} \rightarrow \ket{3}$ transition with Rabi frequency $\Omega_c^{F\left( B\right) }$ and detuning $\Delta_c^{F\left( B\right) }$.
The probe fields (blue arrows) drive the $\ket{1} \rightarrow \ket{3}$  transition with Rabi frequency $\Omega_p^{F\left( B\right) }$ and detuning $\Delta_p^{F\left( B\right) }$.
The superscript $F$ and $B$ denote the forward and backward propagating quantities along $y$ axis, respectively.
Having two-photon resonance for EIT, namely, $\Delta_c^{F\left( B\right) }=\Delta_p^{F\left( B\right) }$, 
the system dynamics is governed by the optical Bloch equation (OBE) \cite{fleischhauer2000, bajcsy2003, Lin2009, Otterbach2010, Kuan2023, Shia2025}
\begin{eqnarray}
\partial_t\rho_{21}  &=&  \frac{i}{2}  \left( \Omega^{F \ast}_c\rho^F_{31} + \Omega^{B \ast}_c\rho^B_{31} \right) ,\label{eq1} \\
\partial_t\rho_{31}^F &=& \frac{i}{2} \Omega_p^F + \frac{i}{2} \Omega_c^F \rho_{21} - \left(\frac{\Gamma}{2}+i\Delta_p^F\right) \rho_{31}^F ,\label{eq2} \\
\partial_t\rho_{31}^B &=& \frac{i}{2} \Omega_p^B + \frac{i}{2} \Omega_c^B \rho_{21} - \left(\frac{\Gamma}{2}+i\Delta_p^B\right) \rho_{31}^B ,\label{eq3}
\end{eqnarray}
\begin{eqnarray}
\frac{1}{c} \partial_t \Omega_{p}^F &+&  \partial_y \Omega_{p}^F = i\eta \rho_{31}^F + \frac{i}{2k_p} \partial_z^2 \Omega_p^F, \label{eq4} \\
\frac{1}{c} \partial_t \Omega_{p}^B &-&  \partial_y \Omega_{p}^B = i\eta \rho_{31}^B + \frac{i}{2k_p} \partial_z^2 \Omega_p^B, \label{eq5}
\end{eqnarray}
where $\rho_{21}$ represents the coherence between $\ket{2}$ and $\ket{1}$. 
$\rho_{31}^{F\left(  B\right) }$ is the coherence between $\ket{3}$ and $\ket{1}$ driven by the forward (backward) probe field $\Omega_{p}^{F\left( B\right) }$.
$\Gamma$ is the spontaneous decay rate of the state $\ket{3}$.
$\eta=\frac{\Gamma\xi}{2L}$ is the light-matter coupling constant, where $\xi $ the optical depth and $L$ the medium length along the $y$-axis.
$k_p = 2\pi/\lambda_p$ is the wavenumber of probe fields, $\lambda_p$ is the probe wavelength, and $c$ is the speed of light in vacuum. For simplicity, we set $\Delta_{p\left( c\right) }^F=\Delta_{p\left( c\right) }^B=\Delta_{p\left( c\right) }$ and two-photon resonance in the following analysis.

We transfer Eqs.~(\ref{eq1}-\ref{eq5}) to a two-dimensional Schr\"odinger equation of the dark-state polarization $\rho_{21}$  carrying effective charge one \cite{Kuan2023, Shia2025}
\begin{equation}\label{eq6}
i\hbar \partial_t \rho_{21} =  \frac{\left( \frac{\hbar}{i} \partial_z - A_z \right)^2}{2M_z} \rho_{21} +\frac{\left( \frac{\hbar}{i} \partial_y - A_y \right)^2}{2M_y} \rho_{21} + U\rho_{21},
\end{equation}
where $\hbar$ is the reduced Planck constant. 
The effective mass $M_{z\left( y\right) }$, the $ {z\left( y\right) } $ component of the synthetic vector potential $A_{z\left( y\right) }$, and the synthetic scalar potential energy $U$ are 
\begin{eqnarray}
M_z &=& \frac{\hbar k_p}{V_F+V_B} , \label{eq7} \\
M_y &=& \frac{\hbar \eta}{2\left(2\Delta_p-i\Gamma\right)\left(V_F+V_B\right)} , \label{eq8} \\
A_z &\approx & \frac{iM_z}{2\eta k_p} \left(  \Omega_c^{F*} \partial_z \Omega_c^F + \Omega_c^{B*} \partial_z \Omega_c^B \right) , \label{eq9} \\
A_y &\approx &  M_y \left( V_B -V_F \right), \label{eq10} \\
U   &\approx &  \frac{i \hbar}{2 \eta} \left(  \Omega_c^{B*} \partial_y \Omega_c^B - \Omega_c^{F*} \partial_y \Omega_c^F \right)   \nonumber\\
&-&\frac{A_z^2}{2M_z}-\frac{A_y^2}{2M_y}+\frac{\hbar\partial_z A_z}{2i M_z}+\frac{\hbar\partial_y A_y}{2i M_y},
 \label{eq11}
\end{eqnarray}
where $V_{F\left( B\right) }=\frac{\left|\Omega_c^{F\left( B\right) }\right|^2}{2\eta}$ denote the forward (backward) EIT group velocity.
We aim to generate a synthetic trapping potential for $\rho_{21}$ by employing biased Gaussian beams as the control fields:
\begin{eqnarray}
\Omega_c^F \left(\vec{r}\right) &=& \Omega \left\lbrace 1 + \alpha G \left(\vec{r}\right) e^{-i \Phi\left(\vec{r}\right)  } \right\rbrace , \label{eq12} \\
\Omega_c^B \left(\vec{r}\right) &=& \Omega \left\lbrace 1 + \alpha G \left(\vec{r}\right) e^{ i \left[  \Phi\left(\vec{r}\right) + 2\phi \right] } \right\rbrace . \label{eq13}
\end{eqnarray}
Here each Gaussian beams is biased by a uniform field $\Omega$. The parameters $\alpha$ represents the fractional contribution of the Gaussian beam component with
the field amplitude $G(\vec{r}) = \frac{1}{\sqrt{1+y^2 / L_r^2}}\exp\left[-\frac{z^2}{ w_0^2\left( 1+y^2 / L_r^2 \right)}\right]$ and the phase $\Phi(\vec{r}) = \frac{z^2}{w_0^2\left( 1+y^2 / L_r^2 \right)} \frac{y}{L_r}$.  
$L_r = \pi w_0^2/\lambda_c$ is the Rayleigh range, $w_0$ is the beam  waist radius, and $\lambda_c$ is the wavelength of  control fields.
$2\phi$ denotes the phase shift between the backward Gaussian beam and the backward uniform field.
In the following, we perform numerical calculations using the two-photon resonance $\Delta_p = \Delta_c$ and the common parameters $\left( \alpha, \Gamma, \Omega,  \lambda_c, \lambda_p, L  \right) = \left(  -0.25,  2\pi \times 6 \text{ MHz}, 3\Gamma,   780 \text{ nm},  780 \text{ nm}, 5 \text{ mm}   \right) $.

We consider the situation $L_r \gg L$ where
Eqs.~(\ref{eq7}-\ref{eq13}) are uniform along $y$ inside the Rayleigh range, and Eq.~\eqref{eq6} can be approximated as
\begin{eqnarray}
i\hbar \partial_t \rho_{21} &\approx & - \frac{\hbar^2}{2M_z} \partial_z^2 \rho_{21} -\frac{\hbar^2}{2M_y} \partial_y^2 \rho_{21} + \frac{i\hbar A_y}{M_y} \partial_y \rho_{21}, \label{eq14} \\
M_z &=& \frac{\hbar k_p \eta }{\Omega^2  \left[     1 + 2\alpha \cos^2\left( \phi\right)   e^{ -\frac{z^2}{w_0^2}} +\alpha^2 e^{ -2\frac{ z^2}{w_0^2}} \right] },    \nonumber\\
M_y &=& \frac{ \eta M_z }{2 \left( 2\Delta_p - i\Gamma \right) k_p  },\nonumber\\
A_y &=& \frac{-2\alpha M_y\Omega^2 \sin^2\left( \phi\right) }{\eta}  e^{ -\frac{z^2}{w_0^2}} . \nonumber
\end{eqnarray} 
We substitute 
$\rho_{21} = \psi_{nm}\left( z\right) \cos\left[ \left( m+\frac{1}{2}\right) k_y y\right]  e^{ \frac{i }{\hbar} A_y y-i\nu_{nm} t} $ and $k_y = 2\pi/L$  into Eq.~\eqref{eq14} and arrive at $- \frac{\hbar^2}{2M_z} \partial_z^2 \psi_{nm} + U_{m} \psi_{nm} = \hbar \nu_{nm} \psi_{nm}$, 
where the effective potential energy reads
\begin{equation}\label{eq15}
U_{m} \approx \left( m+\frac{1}{2}\right) ^2\frac{\hbar^2k_y^2}{2M_y}-\frac{A_y^2}{2M_y}
\end{equation}
Since $U_{m}$ results from the inhomogeneity of $M_y$, we term it as the inhomogeneous mass trap (IMT) for the coherence $\rho_{21}$. 
Fig.\ref{fig1}(c) illustrates the IMT potential $U_{0}/\hbar$ for a negative parameter $\alpha=-0.25$, and Fig.\ref{fig1}(d) shows IMT for a positive  $\alpha=0.25$. In both panels, the real part of IMT Re$\left[ U_{0}/\hbar\right]$ is shown for three different probe detunings: $\Delta_p = -\Gamma$ (red-solid line), $\Delta_p = 0$ (green-solid line), and $\Delta_p = \Gamma$ (blue-solid line). The remaining parameters are set to $\left(  w_0, \phi, \xi\right) = \left(  1 \text{ mm}, 0, 80\right) $ and the longitudinal quantum number $m=0$. 
Given that a convex potential, e.g., the blue solid line in Fig.~\ref{fig1}(c) and the red solid line in Fig.~\ref{fig1}(d), typically generates bound states, one might expect the bound-state condition to be $\alpha \Delta_p < 0$. However, our numerical solutions of the OBE show that centrally confined DSP profiles occur only when $\alpha < 0$.
We find that this apparent discrepancy can be explained by the non-Hermiticity of the IMT. The imaginary part of the IMT Im$\left[ U_{0} \right]$ (black dashed line for all $\Delta_p$) behaves like an incoherent spatial filter that a negative $\alpha$ gives a concave Im$\left[ U_{0}\right]$ (Fig.\ref{fig1}(c)) and leads to stronger attenuation of $\rho_{21}$ outside the Gaussian beam waist than inside, whereas a positive $\alpha$ causes the opposite behavior (Fig.~\ref{fig1}(d)). 
By calculating the trap depth $D_{m}=U_{m}\vert_{ z=\infty} - U_{m}\vert_{ z=0}$, we obtain
\begin{eqnarray}
D_{m} &=& -\frac{\hbar\alpha \left[  \alpha + 2\cos^2(\phi)  \right] \left( 2\Delta_p - i\Gamma\right) \left( 1+ 2 m\right) ^2 k_y^2 \Omega^2 }{4 \eta^2} \nonumber\\ 
&+& \frac{\hbar \left( 2 \Delta_p + i\Gamma \right) \alpha^2\Omega^2 \sin^4(\phi) }{\left( 4 \Delta_p^2 +\Gamma^2 \right) \left[  1+2\alpha \cos^2(\phi)  +\alpha^2 \right]}. \label{eq16}
\end{eqnarray}
Eq.~\eqref{eq16} confirms that  $\alpha < 0$  makes Im$\left[ D_{m} \right] < 0 $ and incoherently results in a centrally confined  profile.
Consequently, the combination of $\alpha < 0$ and $\Delta_p > 0$ will coherently generate a SLP bound state as indicated by the blue line in Fig.\ref{fig1}(c). As $\Omega_c^{F\left( B\right) }$ increases radially along the $z$ axis, SLPs near the center experience weaker attenuation and more dispersion than those at the edges.
In cold-atom media, the use of $\Delta_p^F \neq \Delta_p^B$ suppresses the SLP leakage \cite{Lin2009}. Eq.~\eqref{eq15} can then be generalized for $\phi = 0$ by replacing $2\Delta_p$ with $\Delta_p^F + \Delta_p^B$.
In the following, we focus on the case $\alpha < 0$, which produces the desired potential well and supports bound DSP states.

\begin{figure}[b]
\includegraphics[width=0.48\textwidth]{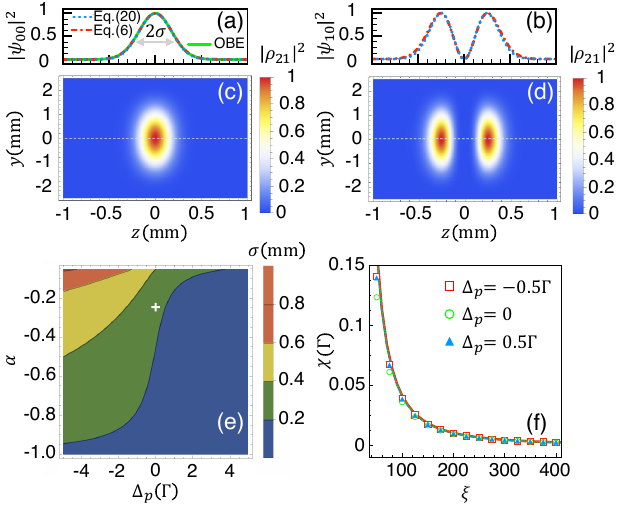}
\caption{\label{fig2}
(a) $\vert \psi_{00} \vert^2$ and (b) $\vert \psi_{10} \vert^2$ are the cross sections along white-dashed lines in (c) ground- and (d) first-excited-state $\vert \rho_{21} \vert^2$, respectively.
Green-solid line is the steady-state  $\vert \rho_{21s}  \vert^2$ from OBE, red-dash-dot lines are the eigenfunction of Eq.~\eqref{eq6}, and blue-dash lines are  Eq.~\eqref{eq20}.
(e) Half width $\sigma$  of  $\vert\psi_{00}\vert^2$.
The white cross pinpoints parameters  $\left( \Delta_p, w_0, \phi, \xi\right) = \left( 0, 1 \text{ mm}, 0, 80\right) $ used in (a-d). (f) Red squares, green circles, and blue triangles are $\chi$ from the numerical solution of  OBE with above parameters except $\Delta_p=-0.5\Gamma$, $0$, and $0.5\Gamma$, respectively. Color solid lines are Eq.~\eqref{eq22}.
}
\end{figure}
We use the Taylor's expansion of $U_m\left( z\right) $ at $z=0$ and obtain the  time-independent Schr\"odinger equation for the one dimensional quantum harmonic oscillator with complex frequency \cite{Jannussis1986}
\begin{equation}\label{eq17}
\frac{-\hbar^2}{2M_z^c} \partial_z^2 \psi_{nm} + \frac{M_z^c \omega_m^2}{2} z^2 \psi_{nm} =  \left( n + \frac{1}{2}\right) \hbar \omega_m  \psi_{nm}. 
\end{equation}
Its eigenvalues $ \left( n + \frac{1}{2}\right) \omega_m = \nu_{nm} - \delta_m$
and  eigenfunctions are
\begin{eqnarray}
\delta_m &=&  \left( m+\frac{1}{2}\right) ^2\frac{\hbar k_y^2}{2M_y^c}-\frac{\left( A_y^{c}\right) ^2}{2\hbar M_y^c}, \label{eq18}\\
\omega_m^2 &=& \left( m+\frac{1}{2}\right) ^2\frac{2\hbar^2 k_y^2 \left( -\alpha\right) \left[ \alpha + \cos^2(\phi)  \right]  }{M_y^c M_z^c w_0^2 \left[  1 + 2\alpha \cos^2(\phi) + \alpha^2 \right]} \nonumber\\
&+& \frac{2 \left( A_y^{c} \right) ^2 \left[ 1 + \alpha \cos^2(\phi) \right] }{ M_y^c M_z^c w_0^2 \left[  1 + 2\alpha \cos^2(\phi)  +\alpha^2 \right]}, \label{eq19}\\
\psi_{nm} &=& c_{nm} H_n\left( \sqrt{\frac{M_z^c \omega_m}{\hbar}} z\right) \exp\left( -\frac{M_z^c \omega_m}{2\hbar} z^2\right),\label{eq20}
\end{eqnarray}
where 
$M_y^c = \frac{\hbar \eta^2 \left( 2\Delta_p + i\Gamma \right)}{2  \Omega^2  \left[  1 + 2\alpha \cos^2(\phi)  +\alpha^2 \right]  \left( 4\Delta_p^2 +\Gamma^2 \right)} = \vert M_y^c\vert e^{i \theta} $ and
$M_z^c = \frac{\hbar k_p \eta}{\Omega^2 \left[  1 + 2\alpha \cos^2(\phi)  +\alpha^2 \right] }$
are the central effective masses, and 
$A_y^{c} = \frac{ -  \hbar \alpha \eta  \left( 2\Delta_p + i\Gamma \right) \sin^2 (\phi) }{  \left[  1 + 2\alpha \cos^2(\phi)  +\alpha^2 \right]  \left( 4\Delta_p^2 +\Gamma^2 \right)}$ is the central vector potential at  $ z = 0 $.   
$\vert M_y^c\vert = \frac{\hbar \eta^2}{2  \Omega^2 \left[  1 + 2\alpha \cos^2(\phi)  +\alpha^2 \right]  \sqrt{4\Delta_p^2+\Gamma^2}}$ the mass module,  $\theta = \cos^{-1}\left( \frac{2\Delta_p}{\sqrt{4\Delta_p^2 + \Gamma^2}}\right) $ the mass angle, and $c_{nm}$ is the normalization constant.
Eq.~(\ref{eq14}-\ref{eq16}) suggest the physical picture that two counter-propagating biased Gaussian fields $\Omega_c^{F\left( B\right) }$ on one hand constitute a cavity along the $y$ direction, and, on the other hand,  generate a harmonic trap along the $z$ direction.
Fig.\ref{fig2} compares the analytic solution Eq.~\eqref{eq20} (blue-dotted line) with results from two numerical methods \cite{Kuan2023, Shia2025}: steady-state solution of OBE, i.e., Eqs.(\ref{eq1}-\ref{eq5}),  (green-solid line), and the eigenfunction of Eq.\eqref{eq6} (red-dashed-dotted line). 
As higher energy states attenuate faster than the ground state, the steady-state solution of the OBE is expected to converge to the ground state.
The panels Fig.\ref{fig2}(a) and (b) illustrate the probability densities $\vert \psi_{00} \vert^2$ and $\vert \psi_{10} \vert^2$, respectively. These probability density plots are taken as cross-sections along the white dashed lines shown in Fig.\ref{fig2}(c) (ground state) and Fig.\ref{fig2}(d) (first excited state) from the numerical diagonalization of Eq.\eqref{eq6}.
Both figures confirm a good agreement between the analytic and numerical solutions with  parameters  $\left( \Delta_p, w_0, \phi, \xi \right) = \left( 0, 1 \text{ mm}, 0, 80 \right) $.
Moreover, we derive the ground-state probability density $\vert\psi_{00}\vert^2 \propto \exp\left( -z^2 / \sigma^2 - \chi t\right) $ for $\phi = 0$, where the half width $\sigma$ and the decay rate $\chi$ are 
\begin{eqnarray}
&\sigma =  \sqrt{ \frac{w_0 }{k_y} } \left[ \frac{2\eta\left( 1+\alpha\right) }{-\alpha k_p \left( 2\Delta_p + \sqrt{\Gamma^2 + 4 \Delta_p^2 } \right) }\right] ^{\frac{1}{4}}, \label{eq21}\\
&\chi =  \frac{\Gamma k_y^2 \Omega^2 \left( 1+\alpha\right) ^2 }{2\eta^2} + \frac{k_y \Gamma\Omega^2\sqrt{-\alpha\left( 1+\alpha\right) ^3}}{w_0  \sqrt{2 k_p \eta^3 \left( 2\Delta_p + \sqrt{\Gamma^2 + 4 \Delta_p^2 } \right)}}.\label{eq22}
\end{eqnarray} 
Based on Fig.\ref{fig1}(c), $\Delta_p > 0$ generates a deeper IMT  than $\Delta_p \leq 0$ does. This deeper well confines  $\rho_{21}$ more tightly, as illustrated in Fig.\ref{fig2}(e). 
As a result, $\Delta_p > 0$ leads to a narrower spatial spread $\sigma$ than $\Delta_p \leq 0$ does. 
Fig.\ref{fig2}(f) demonstrates the excellent agreement between the analytical Eq.~\eqref{eq22} (solid lines) and the $\chi$ values obtained from the numerical solution of OBE with $\Delta_p=-0.5\Gamma$ (red squares), $0$ (green circles), and $0.5\Gamma$ (blue triangles). This comparison utilized the same parameters employed in Fig.\ref{fig2}(a-d), where the numerical $\chi$ was determined by calculating the attenuation of the initial ground state.
Analysis of Eq.~\eqref{eq22} shows that the first term is dominant, and the dependence on $\Delta_p$ is negligible.

\begin{figure}[b]
\includegraphics[width=0.48\textwidth]{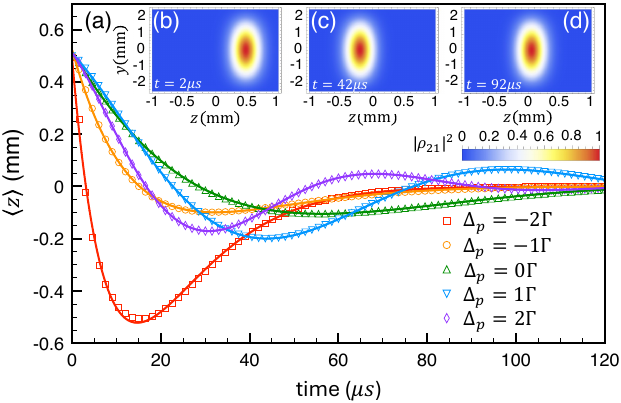}
\caption{\label{fig3}
(a) Coherent state oscillation of the expectation value $\left\langle z\right\rangle $ for $\Delta_p=-2\Gamma$ (red squares), $\Delta_p=-1\Gamma$ (orange circles),  $\Delta_p=0\Gamma$ (green triangles),  $\Delta_p=1\Gamma$ (blue-inverted  triangles), and $\Delta_p=2\Gamma$ (purple  diamonds). The insets depict the snapshot of $\vert \rho_{21}\left( z,y\right)  \vert^2$ for $\left( \Delta_p, w_0, \phi, \xi\right) = \left(  1\Gamma, 1.5 \text{ mm}, 0, 80\right) $ at (b) $t=2\mu s$, (c) $t=42\mu s$, and (d) $t=92\mu s$.
}
\end{figure}
We next examine coherent effects by analyzing $\rho_{21}$ in a coherent state and demonstrate that the IMT goes beyond a simple spatial filtering mechanism. 
We numerically solve OBE with the parameters $\left(  w_0, \phi, \xi \right) = \left(  1.5 \text{ mm}, 0, 80 \right) $ to investigate the  dynamics of the coherent-state $\rho_{21} = c_{00}\exp\left[ -\frac{M_z^c \omega_m}{2\hbar} \left( z-z_0\right)^2\right] $. Here $z_0$ represents the peak position which is intentionally shifted from the IMT center. This initial wavepacket is prepared and deployed via the EIT light storage and retrieval technique \cite{Phillips2001, Kocharovskaya2001, Chen2013, Hsiao2018, Wang2019}.
We then calculate the subsequent expectation value of the wavepacket's position
$\langle z\left( t\right) \rangle = \left[ \int z \vert \rho_{21}\left( \vec{r}, t\right) \vert^2 d^2\vec{r}\right] /\int  \vert \rho_{21}\left( \vec{r}, t\right) \vert^2 d^2\vec{r}$.
Fig.\ref{fig3}(a) presents the temporal evolution of $\langle z(t) \rangle$ for different probe detunings $\Delta_p=-2\Gamma$ (red squares), $-1\Gamma$ (orange circles), $0$ (green triangles), $1\Gamma$ (blue-inverted triangles), and $2\Gamma$ (purple diamonds). 
We observe the typical damped oscillation in the wavepacket's mean position $\langle z(t) \rangle$.
For example, snapshots of $\vert \rho_{21}\left( \vec{r}, t\right) \vert^2$ with $\Delta_p=1\Gamma$ are shown in Fig.~\ref{fig3}(b) at $t=2~\mu s$, Fig.~\ref{fig3}(c) at $t=42~\mu s$, and Fig.~\ref{fig3}(d) at $t=92~\mu s$. These snapshots show that the entire wave packet oscillates back and forth.
This behavior is quantitatively analyzed by fitting the numerical results to the standard formula for a damped harmonic oscillator $e^{-\kappa t}\cos\left( ft\right) $. Each solid line in Fig.\ref{fig3}(a) shows the optimal fitting curve with the fitting parameters  
$\left( f, \kappa \right) = \left( 21.1 \text{kHz}, 84.7 \text{kHz}\right)$ for $\Delta_p=-2\Gamma$, 
$\left( f, \kappa \right) = \left( 27.4 \text{kHz}, 54.9 \text{kHz}\right)$ for $\Delta_p=-1\Gamma$, 
$\left( f, \kappa \right) = \left( 29 \text{kHz},   28 \text{kHz}\right)$ for $\Delta_p=0$, 
$\left( f, \kappa \right) = \left( 58.1 \text{kHz}, 20.8 \text{kHz}\right)$ for $\Delta_p=1\Gamma$, and
$\left( f, \kappa \right) = \left( 84.1 \text{kHz}, 33.7 \text{kHz}\right)$ for $\Delta_p=2\Gamma$.
The fitted oscillation frequency $f$ shows a good agreement with the theoretical prediction of the real part of the trap frequency Eq.~\eqref{eq19} for $\Delta_p > 0$ giving 
Re$\left[ \omega_0\right] = 66$   kHz for  $\Delta_p= 1\Gamma$, and 
Re$\left[ \omega_0\right] = 91.6$ kHz for  $\Delta_p= 2\Gamma$.
The underdamped oscillation occurs only for $\Delta_p > 0$, and the oscillation frequency increases with increasing $\Delta_p$. 
Moreover, for $\Delta_p \leq 0$, the wavepacket crosses $z=0$ only once before returning to the origin.
Both behaviors confirm the picture in Fig.~\ref{fig1}(c), namely, the combination of $\alpha < 0$ and $\Delta_p > 0$ produces a deeper IMT than the case with $\Delta_p < 0$. 
Figs.~\ref{fig2} and \ref{fig3} show excellent agreement between numerical and theoretical results, as well as the remarkable controllability of attenuation and coherent dynamics in the non-Hermitian IMT.

\begin{figure}[b]
\includegraphics[width=0.48\textwidth]{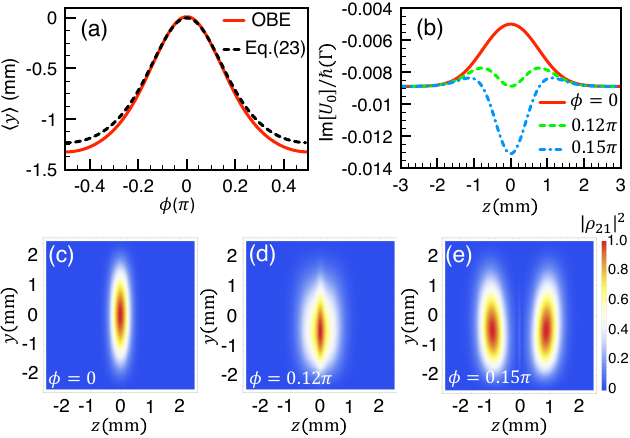}
\caption{\label{fig4}
(a) $\phi$-dependent expectation value $\langle y \rangle$ at $z=0$. The red-solid line shows the numerical solution of the OBE, and the black-dashed line represents Eq.~\eqref{eq23}.
(b) Im$\left[ U_0\right] /\hbar$ for $\phi = 0$ (red-solid line),  $\phi = \phi_c = 0.12\pi$ (green-dashed line), and  $\phi = 0.15\pi$ (blue-dashed-dotted line).
(c) $\vert \rho_{21}(z,y) \vert^2$ for $\phi = 0$, (d)  $\phi = 0.12\pi$, and (e)  $\phi = 0.15\pi$.
Other parameters are  $\left(  \Delta_p, w_0,   \xi\right) = \left(   1\Gamma, 1.5 \text{ mm},   200\right) $. 
}
\end{figure}
%
%
%
We now study the effect of the Gaussian beam phase shift $\phi $ and calculate the ground-state position expectation value $\langle y\rangle = \left( \int y \vert \rho_{21}\vert^2 dy\right) / \int  \vert \rho_{21}\vert^2 dy$, where 
the the ground-state $\vert\rho_{21}\vert^2\propto  \exp  \left\lbrace -\frac{2}{\hbar} \text{Im} \left[ A_y\right] y \right\rbrace $.
The exponent  reveals a key feature that  $A_y$ can push or pull DSPs along the $y$ direction. For instance, when $ \frac{2}{\hbar} \text{Im} \left[ A_y\right]  > 0$, the vector potential introduces an asymmetry and pushes the peak of the wave function towards the region  $y < 0$.
By evaluating $\langle y\rangle$, we obtain the analytical expression for the $\phi$-dependent displacement
\begin{equation}
\left\langle y \right\rangle = \frac{3k_1^2+k_y^2}{k_1\left(k_1^2+k_y^2\right)} 
- \frac{\pi}{k_y} \coth\left(\pi\frac{k_1}{k_y}\right), \label{eq23}
\end{equation}		
where $k_1 = \frac{2}{\hbar} \text{Im} \left[ A_y\right]$.
Fig.\ref{fig4}(a) presents the comparison between  Eq.\eqref{eq23} (black-dashed line) and the numerical $\langle y\rangle$  from OBE (red-solid line) with the  parameters $\left( \Delta_p, w_0,  \xi\right) = \left(  1\Gamma, 1.5 \text{ mm},  200\right) $ at $z=0$.
The good agreement between the analytical result and the numerical solution confirms the predicted $\phi$-induced negative displacement of the ground-state wavepacket.
Fig.\ref{fig4}(c), (d), and (e) display the spatial distribution of the steady-state  $\vert \rho_{21}\vert^2$ for $\phi = 0$, $0.12\pi$, and  $0.15\pi$, respectively. We note that as  $\phi$ is increased, the wavepacket not only undergoes a backward displacement but also exhibits a distinct splitting when $\phi > 0.12\pi$. 
The $\phi$ induced splitting of $\rho_{21}$ can be attributed to  the non-uniform attenuation. Fig.\ref{fig4}(b) illustrates the imaginary part of IMT for $\phi = 0$ (red-solid line),  $\phi =  0.12\pi$ (green-dashed line), and  $\phi = 0.15\pi$ (blue-dashed-dotted line). 
For $\phi = 0$, the concave profile of Im$\left[ U_0\right] $ leads to weaker attenuation of $\rho_{21}$ inside the Gaussian beam waist than outside. As $\phi$ increases, Im$\left[ U_0\right] $ becomes convex, causing greater attenuation inside the beam waist than outside. This spatially varying attenuation results in the observed splitting of $\rho_{21}$.
We determine the critical phase $\phi_c$ at which the splitting emerges by solving Im$\left[ D_0\right] = 0 $, resulting in a well-defined threshold condition
\begin{equation}\label{eq24}
\phi_c = \tan ^{-1}\sqrt{\frac{  -\left(  2 + \alpha \right) \left(  1+\alpha \right) ^2  }{1+ \alpha \left(  1+ \alpha \right) ^2-\sqrt{1-\frac{4 \alpha \left( 2+\alpha \right) \left( 1+\alpha \right) ^2   \eta ^2}{k_y^2 \left(\Gamma ^2+4 \Delta_p^2\right)}}}}.
\end{equation}
Eq.~\eqref{eq24} yields the exact numerical value $\phi_c = 0.12\pi$ for the parameters used in Fig.~\ref{fig4}, in agreement with the observed results.

In conclusion, we have demonstrated a way to generate IMT for DSPs invoking two counter-propagating biased Gaussian beams under the guidance of the  Schr\"odinger-like Eq.\eqref{eq6}. 
This IMT configuration allows for highly tunable engineering of the DSPs' properties, including their attenuation, spatial profile, and coherent dynamics. This control is achieved by manipulating parameters, e.g., the laser detuning, the coupling field strength, the Gaussian beam phase, and the optical depth. This confirms the present EIT system as a useful platform for controlling DSPs.

This work is supported by National Science and Technology Council, Taiwan (112-2112-M-007-020-MY3 \& 113-2628-M-008-006-MY3).

\bibliography{20260316_LGEIT}
\end{document}